\documentclass[sigconf,nonacm]{acmart}

\setcopyright{none}
\copyrightyear{2025}
\acmYear{2025}

\usepackage{balance}

\usepackage{hyperref}
\makeatletter
\g@addto@macro{\UrlBreaks}{\UrlOrds}
\makeatother

\usepackage{hyperref}
\usepackage{amsmath}
\usepackage{amsfonts}
\usepackage{tcolorbox}
\usepackage{longtable}
\usepackage{listings}

\tcbset{
  mybox/.style={
    colframe=black,
    colback=white,
    sharp corners,
    boxrule=1pt,
    leftrule=3pt,
    right=0pt,
    top=0pt,
    bottom=0pt,
    boxsep=5pt,
    arc=5pt,
  }
}

\lstdefinestyle{SQLstyle}{
  language=SQL,
    basicstyle=\footnotesize\ttfamily, %
  keywordstyle=\bfseries,
  commentstyle=\itshape,
  showstringspaces=false,
  numbers=left,
  numberstyle=\tiny,
  frame=single,
  breaklines=true,
  captionpos=b,
  escapechar=!,
escapeinside={(*@}{@*)}, %
}
\lstdefinestyle{JSONstyle}{
  language=Python,
    basicstyle=\footnotesize\ttfamily, %
  keywordstyle=\color{blue},
  stringstyle=\color{orange},
  commentstyle=\color{gray},
  showstringspaces=false,
  numbers=left,
  numberstyle=\tiny,
  frame=single,
  breaklines=true,
  captionpos=b
}

\usepackage{bbm}
\usepackage{dsfont}

\usepackage{algorithm}[h]
\usepackage[noend]{algpseudocode}

\usepackage{paralist}
\usepackage{booktabs} %
\usepackage{multirow}
\usepackage{enumitem}

\usepackage{subcaption}
\usepackage{epstopdf}
\usepackage{graphics}
\DeclareGraphicsExtensions{.eps,.gz,.eps,.pdf,.png,.jpg}
\graphicspath{{./}{./fig/}}

\usepackage{graphicx}

\newcommand{\sstitle}[1]{\smallskip\noindent\textbf{#1.\/}}

\def\Snospace~{\S{}}

\makeatletter
\newcommand{\removelatexerror}{\let\@latex@error\@gobble}
\makeatother

\usepackage{xspace}
\newcommand{\toolname}{\texttt{MATS}\xspace}

\begin{document}

\title[Multi-agent Text2SQL with Small Language Models]{A Multi-agent Text2SQL Framework using Small Language Models and Execution Feedback}

\author[Thanh Dat Hoang et al.]{Thanh Dat Hoang$^1$,
Thanh Trung Huynh$^2$, Matthias Weidlich$^3$, Thanh Tam Nguyen$^1$, Tong
Chen$^4$, Hongzhi Yin$^4$, Quoc Viet Hung Nguyen$^1$}
\affiliation{
\institution{$^1$Griffith University (Australia), $^2$VinUniversity (Vietnam),\\ $^3$Humboldt-Universitat zu Berlin (Germany), $^4$The University of
Queensland, Australia}
\country{\vspace{.8em}}
}

\renewcommand{\shortauthors}{Thanh Dat Hoang et al.}

\begin{abstract}
Text2SQL, the task of generating SQL queries from natural language text, is a critical challenge in data engineering. Recently, Large Language Models (LLMs) have demonstrated superior performance for this task due to their advanced comprehension and generation capabilities. However, privacy and cost considerations prevent companies from using Text2SQL solutions based on external LLMs offered as a service. Rather, small LLMs (SLMs) that are openly available and can hosted in-house are adopted. These SLMs, in turn, lack the generalization capabilities of larger LLMs, which impairs their effectiveness for complex tasks such as Text2SQL. To address these limitations, we propose \toolname, a novel Text2SQL framework designed specifically for SLMs. \toolname uses a multi-agent mechanism that assigns specialized roles to auxiliary agents, reducing individual workloads and fostering interaction. A training scheme based on reinforcement learning aligns these agents using feedback obtained during execution, thereby maintaining competitive performance despite a limited LLM size. Evaluation results using on benchmark datasets show that \toolname, deployed on a single-GPU server, yields accuracy that are on-par with large-scale LLMs when using significantly fewer parameters. Our source code and data are available at \url{https://github.com/thanhdath/mats-sql}.

\end{abstract}



\maketitle

\section{Introduction}
\label{sec:intro}

Text2SQL, the task of translating natural language into SQL queries, is a
long-standing research challenge~\cite{zhong2017seq2sql,xu2017sqlnet,
yu2018typesql,yu2018syntaxsqlnet}. While an increasing complexity of user
queries and database schemas contribute to the task's
difficulty~\cite{rokis2022challenges}, recent solutions based on Large
Language Models (LLMs) achieved notable results for
Text2SQL~\cite{pourreza2023dinsql,sun2023sqlpalm,dong2023c3,liu2023comprehensive}.
Text2SQL approaches promise to
enable non-experts to query databases using a natural
language interface~\cite{touvron2023llama,papicchio2025qatch}.
Most existing solutions for this task, however, rely on external LLMs
offered as a
service~\cite{naveed2025comprehensive,schellaert2025analysing}, primarily variants of OpenAI GPT, to generate SQL queries. These
approaches combine user queries and schema representations with
instructional text to generate SQL
queries~\cite{roziere2023code,li2023starcoder}. However, the use of
external LLM services comes with drawbacks. Privacy concerns arise
when sensitive data, such as database schemas or query logs, are shared
with third-party platforms, potentially violating confidentiality
or exposing data to security risks. Also, the recurring costs of
such services can become a substantial financial burden, especially
for small organizations.

To be independent of external LLM services, recent approaches for Text2SQL
fine-tune open-source LLMs with instructional
data~\cite{kumar2022finetuning,raffel2020exploring,dong2024abilities}.
Fine-tuning improves the task-specific model performance, but requires
significant computational resources and
technical expertise. In addition, even with models as large as 15 billion
parameters, the accuracy obtained using open-source LLMs is
considerably lower than the one achieved with
external LLM services~\cite{lan2023unite,li2024codes}.
Striving for a cost-effective solution, one may adopt small Large Language
Models (SLMs) with generally smaller numbers of parameters (typically
100M-5B)~\cite{lu2024small}. Such models are optimized to run on a
single-node server that features a single GPU. While the use of SLMs
enlarges the efficiency and, hence, applicability of a Text2SQL solution,
one faces challenges in terms of model effectiveness. SLMs often struggle
with tasks that require deep reasoning or
understanding of complex contexts, such as required for
Text2SQL~\cite{lu2024small}.
Their limited capacity makes it challenging to maintain relationships
between tables or fields, especially with large schemas, and they
frequently
miss nuances in natural language inputs, leading to syntax or semantic
errors in SQL generation.

In this paper, we follow the idea of using SLMs for Text2SQL and propose the
\textbf{M}ulti-\textbf{A}gent \textbf{T}ext2\textbf{S}QL (\toolname) framework to
operationalize it.  \toolname employs a multi-agent
mechanism~\cite{qian2024chatdev, islam-etal-2024-mapcoder, ijcai2024p890} to
decompose the Text2SQL task into sub-tasks, each handled by a specialized
agent: a schema investigator filters irrelevant schema elements and
retrieves relevant column values; a query planner generates multiple SQL
queries step-by-step; a validator evaluates SQL outputs using database
responses; a fix agent  refines SQL based on validator feedback; and a
selection agent, at the end of the pipeline, selects the best SQL query from
the final candidates. To enhance the collaboration between the agents, we
design a collaborative training scheme, coined Reinforcement Learning with
Execution Feedback (RLEF). Unlike traditional Reinforcement Learning from
Human Feedback~\cite{ouyang2022training}, RLEF generates multiple responses
using automated database feedback, avoiding the need for costly
human-labeled data.

\begin{figure}[!h]
\vspace{-0.5em}
	\centering
	\includegraphics[width=0.95\linewidth]{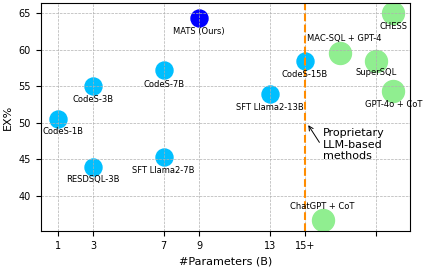}
	\vspace{-1em}
	\caption{Execution accuracy (EX\%) of \toolname vs. other
		approaches on BIRD dev. Methods using open-source models and
		proprietary LLMs are separated by the dashed line.
	}
	\label{fig:motivating}
	\vspace{-.5em}
\end{figure}

The divide-and-conquer strategy realized in \toolname is beneficial in terms of
efficiency
and effectiveness. Due to the specialization of agents and
their focus on a single sub-task, the generalization capabilities of SLMs,
which can be managed efficiently, are sufficient to yield high accuracy. At
the same time, the integration of the agents using reinforcement learning
enables our framework to effectively handle complex user queries and
large-scale datasets. Furthermore, our framework facilitates the adaptation of open-source SLMs, thereby supporting wider applicability on resource-constrained devices and under restricted budgets.

We summarize the contributions of our paper as follows:

\begin{compactitem}%
\item We propose a novel multi-agent framework in which specialized
agents rely on SLMs to collaboratively solve Text2SQL tasks. The framework
defines sub-tasks for element filtering, query planing, validation of query
results, refinement of queries, and query selection.

\item We introduce Reinforcement Learning with Execution Feedback (RLEF) as
a mechanism to enable SLMs agents to collaborate during training,
significantly improving their performance in Text2SQL tasks.
It relies on recent advancements for preference
optimization~\cite{hong-etal-2024-orpo}
and instantiates them based on a sampling scheme for appropriate responses.

\item We create a comprehensive dataset tailored for training SLMs agents by extending the Spider and BIRD datasets through manual labeling, few-shot prompting, and fine-tuning, ensuring high-quality examples for robust learning.

\item We evaluate \toolname in comprehensive experiments and observe that it achieves
results that are on-par with large-scale LLMs, such as GPT-4o +
CoT~\cite{li2023can} and CHESS~\cite{talaei2024chess}, while relying on
significantly smaller models. 

\item Our source code and data are available at \url{https://github.com/thanhdath/mats-sql}.

\end{compactitem}

\autoref{fig:motivating} illustrates the main
insight from our experiments in terms of the relation of the execution
accuracy and the total size of the model in terms of its parameters. With
total model size of 9B, \toolname is optimized for resource-constrained
environments. Our experimental results show that MATS enables efficient inference without
sacrificing performance, which renders it
well-suited for cost-sensitive deployments.

In the remainder of the paper, \autoref{sec:model_problem} formulates the
addressed problem. \autoref{sec:multi-agents} outlines the key components of
the \toolname framework and their instantiation.
\autoref{sec:rlef} introduces our approach to Reinforcement Learning with
Execution Feedback (RLEF).
Evaluation results are presented in \autoref{sec:exp}, before we review our
contributions in the light of
related work in \autoref{sec:related} and conclude the paper in
\autoref{sec:con}.

\section{Model and Requirements}
\label{sec:model_problem}

We first characterize the problem addressed in this work
(\autoref{sec:problem}), before elaborating on requirements for
solutions to it (\autoref{sec:design_principles}).

\subsection{Problem Formulation}
\label{sec:problem}

Text2SQL addresses the task of generating an SQL query $\mathcal{Y}$ that
corresponds to a given natural language question $q$. This query
is constructed based on a database schema ${S}$ and, optionally, an
external knowledge base ${K}$. The database schema ${S}$ is
defined by
a set of tables $\left\{{T}_1, {T}_2, \ldots,
{T}_m\right\}$,
a set of columns
$\left\{{C}_1, {C}_2, \ldots, {C}_n\right\}$, and
a set of foreign key relations $\left\{{R}_1, {R}_2, \ldots, {R}_k\right\}$.
The optional external
knowledge base ${K}$ provides context for the schema, aiding in
generating more accurate SQL in ambiguous situations.

\noindent
Mathematically, the Text2SQL task is formulated as:
$$
\mathcal{Y}=f(q, {S}, {K} \mid \boldsymbol{\theta}),
$$
where the function $f(\cdot \mid \boldsymbol{\theta})$ represents a
generative model (e.g., a neural network) with learnable parameters
$\boldsymbol{\theta}$.

\subsection{Requirements}
\label{sec:design_principles}

We argue that any SLMs-based solution for Text2SQL shall address the
following requirements:

\smallskip
\textit{(R1) Large Database Schema.}
A Text2SQL solution shall handle large database schemas. This is
challenging as the sheer
number of tables and columns can exceed the model's context length,
impairing comprehension. Real-world schemas
often include overlapping column names and extensive metadata, further
complicating the respective task. For example, the BIRD
dataset features databases with up to 65 tables and 455 columns, increasing
the likelihood of errors in schema linking and SQL query
generation~\cite{li2024dawn}.

\smallskip
\textit{(R2) Ambiguous Column Names and Values.}
A Text2SQL solution shall cope with the ambiguity in column names and
values, especially when multiple columns share similar meanings or
overlapping values. For example, names of organizations may appear in
different roles, and hence, as different columns in database.
Correctly linking queries to columns is then challenging and increases the
risk of incorrect SQL generation.

\smallskip
\textit{(R3) Weak Reasoning Capability of SLMs.}
A Text2SQL solution based on SLMs needs to address the limited reasoning
capabilities of the respective models.
Specifically, chain-of-thought prompting, which enhances
reasoning in large models, is less effective for SLMs and can even produce
fluent, yet illogical reasoning outputs~\cite{wei2022chain}. In Text2SQL
tasks, this limitation becomes particularly important and SLMs have been
observed to frequently generate
inaccurate SQL queries for complex database schemas~\cite{li2024dawn}.

\smallskip
\textit{(R4) Low Instruction Following Capability.}
SLMs struggle with instruction following due to their limited parameter
size~\cite{murthy2024evaluating}, which needs to be incorporated in Text2SQL
solutions using these models. SLMs tend to overfit to specific training
formats and typically lack exposure to diverse instruction-tuning datasets,
such as InFoBench~\cite{qin-etal-2024-infobench} or
IFEval~\cite{zhou2023instruction}. Hence, SLMs are limited in their
generalization to new or varied instruction types, as well as instructions
that involve dependencies and sequential logic.

\section{The \toolname Framework}
\label{sec:multi-agents}

\begin{figure*}[!h]
	\centering
	\includegraphics[width=\linewidth]{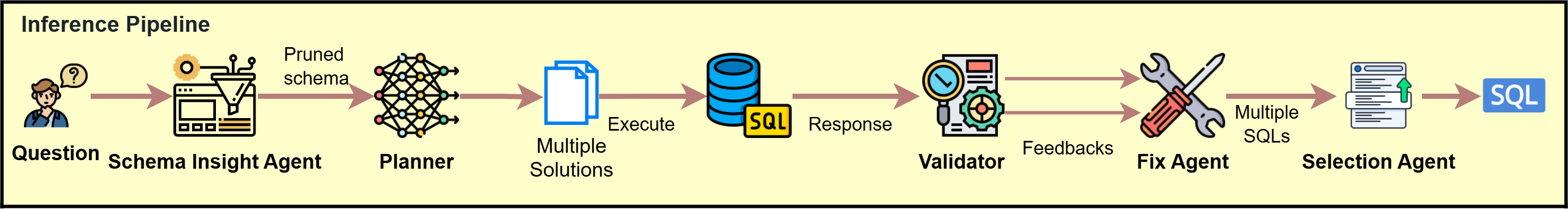}
	\caption{Overview of our multi-agent framework for Text2SQL.}
	\label{fig:framework}
	\vspace{-0.2em}
\end{figure*}

This section presents our \textbf{M}ulti-\textbf{A}gent
\textbf{T}ext2\textbf{S}QL (\toolname) framework. As illustrated in
\autoref{fig:framework}, it adopts the paradigm of multi-agent
collaboration, i.e., a splits the Text2SQL task into sub-tasks that are
handled by individual agents.

Given a user query and database schema, the
process
begins with the \textbf{Schema Insight Agent}, which
extracts relevant tables and columns, even when the query does not exactly
match stored values. As such, this agent explicitly addresses the
requirements of handling large database schemas (R1) and ambiguous column
names and values (R2).
Next, the \textbf{Planner
Agent} decomposes the reasoning process into a chain
of thoughts. It addresses the weak reasoning capabilities (R3) of SLMs by
generating SQL candidates.
The \textbf{Validator Agent}
then evaluates these candidates and their
execution results, identifying potential errors. Any detected issues are
refined by the \textbf{Fix Agent}. Finally, the
\textbf{Selection Agent} chooses the best SQL
query based on execution responses.

In the remainder of this section, we elaborate on the realization of the
individual agents (\autoref{sec:schema} - \autoref{sec:selection}), before
turning to the creation of training data for fine-tuning
(\autoref{sec:data_creation_mechan}), which also caters for the weak
reasoning capabilities (R3) of SLMs.

\subsection{Schema Insight Agent}
\label{sec:schema}
Given a question posed by a user as input, the Schema Insight Agent filters
out irrelevant
schema
elements and retrieves relevant column values. To this end, it includes two
components: Schema Filtering and Value Matching.
Schema Filtering eliminates tables and columns
that are unlikely to contribute to generating the correct SQL query. In our
framework, we adopt CodeS~\cite{li2024codes} for this purpose. It uses a
bidirectional encoder to rank tables and columns, and discards those with
low relevance to the given user question.
Value Matching leverages the BM-25 algorithm~\cite{askari2023injecting} to
identify column
values that closely align with the input query. This functionality is
essential for selecting the appropriate columns for the generation of
accurate SQL queries.
Specifically, given an input question~$q$, for each column \( c_i \) in
the pruned schema \( C = \{ c_1, c_2, \dots, c_n \} \) that is obtained by
Schema Filtering, we retrieve a set of
candidate values \( V_{c_i} = \{ v_{i1}, v_{i2}, \dots, v_{im} \} \). We
compute the BM25 relevance score between the question and each
value:
\[
\text{Score}(q, v_{ij}) = \text{BM25}(q, v_{ij}).
\]
Then, we select the $k$ values with highest scores for each
column:
\[
V_{c_i}^* = \operatorname{Top}_k\left( V_{c_i}, \text{Score}(q, V_{c_i}) \right),
\]
where typically \( k = 2 \).
If no values yield a positive BM25 score, we
select a representative example value from \( V_{c_i} \). The selected
values \( V_{c_i}^* \) are then incorporated into the database schema
prompt, providing the model with contextual cues for SQL generation.

\subsection{Planner Agent}
\label{sec:planner}
The Planner Agent generates SQL queries by decomposing the reasoning process
into small, step-by-step operations, enabling SLMs to construct accurate and
well-structured queries. As the central component of the system, this agent
is responsible for translating user questions into SQL queries to fulfill
the given task. %

\begin{figure}[!h]
	\centering
	\includegraphics[width=0.99\linewidth]{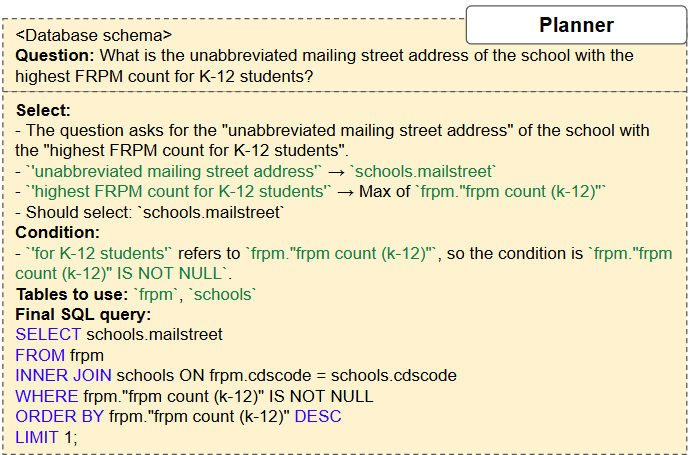}
	\vspace{-1em}
	\caption{Example thought process of the Planner.}
	\label{fig:planner_thought}
	\vspace{-1em}
\end{figure}

We manually design a reasoning process based on few-shot examples that
consists of three steps: 1) identifying the selection goal, 2) analyzing
conditions for the WHERE clause, and 3) determining the necessary tables for
the FROM and JOIN clauses. This systematic approach guides the model
through the process of deciding what to select, which conditions to apply,
and, thus, which tables to use. An example of the thought process is
illustrated in \autoref{fig:planner_thought}.

Let \( x \) represent a data sample containing a question, a database
schema, and, optionally, an external knowledge base. The Planner Agent
generates a plan \( p \) and a SQL query \( s \) as follows:
\begin{equation}
p \leftarrow \pi_p(x)
\end{equation}
\begin{equation}
s \leftarrow \pi_p(x, p)
\end{equation}

In our framework, the Planner produces \( K \) SQL queries: one using greedy
decoding and \( K-1 \) using multinomial decoding with a temperature \(
\mathcal{T} \):
\begin{equation}
S = \{s_1, s_2, \dots, s_K\} \leftarrow \pi_p(x, p, \mathcal{T}).
\end{equation}

\subsection{Validator Agent}
\label{sec:validator}

The Validator Agent re-evaluates the generated SQL query based on the
response received from the database to identify errors. As such, the
Validator not only verifies the Planner's reasoning but also detects issues
that require modifications, e.g., related to syntax errors in the query or
queries that yield empty responses.

While the Validator is similar to the Planner in terms of the goal to
validate selection goals, to determine the relevant tables, and to analyze
the conditions of a selection, this redundancy is crucial, as LLMs often
fail on the first attempt on a task~\cite{madaan2024self}. The Validator
uses the responses from query evaluation to detect discrepancies, to refine
the SQL queries, and to provide targeted feedback when queries fail.
Specifically, we employ two specialized validator agents.

\sstitle{Validator Selection}
The validator assesses that the generated SQL query accurately selects the
correct columns based on the intent of the question. LLM-generated SQL
queries often contain mistakes, such as the
selection of incorrect or unnecessary columns, the omission of necessary
ones, or the arrangement of columns in the wrong order.

\begin{figure}[!h]
	\vspace{-1em}
	\centering
	\includegraphics[width=0.99\linewidth]{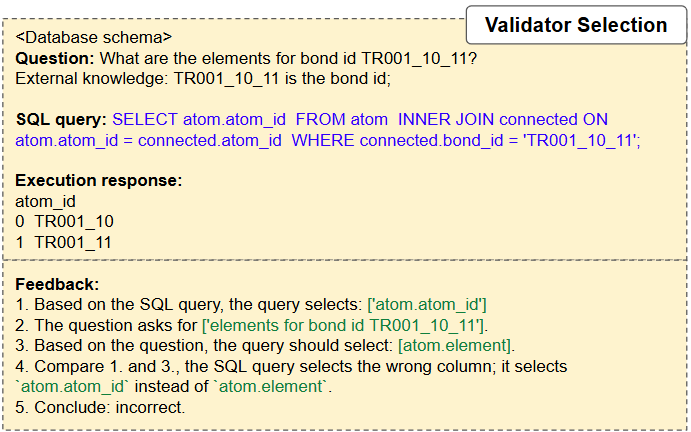}
	\vspace{-1em}
	\caption{Example feedback of validator selection.}
	\label{fig:validator_selection}
	\vspace{-1em}
\end{figure}

We overcome these issues based on a dedicated thought process for the
validator for the selection, as it is illustrated in
\autoref{fig:validator_selection}. First, the validator checks the selected
columns of the input SQL query. Next, it performs phrase extraction on the
input question and maps it to the intended columns that need to be selected.
Finally, the validator compares the selected columns with the intended ones
and flags potential issues.

This validator considers only on queries that do not include certain
operations such as \texttt{min}, \texttt{max}, \texttt{count}, \texttt{avg},
\texttt{sum}, \texttt{divide}, or \texttt{case when}. The reason for this
restriction is that multiple queries selecting different columns may still
produce the same correct result, which could otherwise lead to misjudgment
by the validator. At the end of the validation process, the validator
determines whether the SQL query is correct or not.

\sstitle{Validator Condition}
This validator aims at identifying mistakes related to logical conditions in
SQL queries. In an SQL query, conditions can be used in the \texttt{WHERE}
clause or the \texttt{SELECT} clause (e.g., queries using \texttt{CASE WHEN}
or \texttt{IF} statements).

\begin{figure}[!h]
	\centering
	\includegraphics[width=0.99\linewidth]{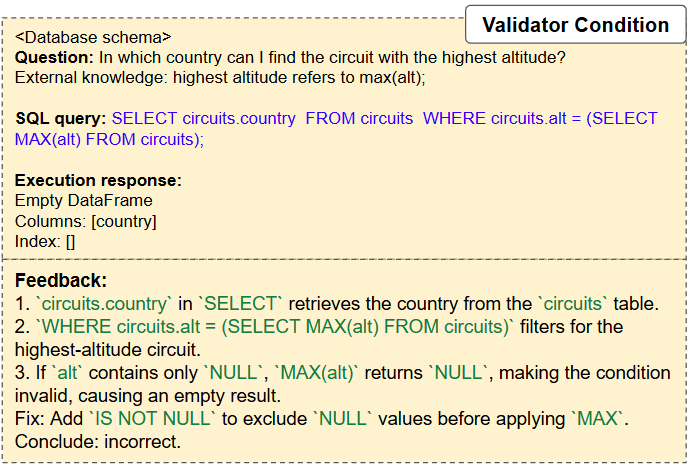}
	\vspace{-1em}
	\caption{Example feedback of validator condition.}
	\label{fig:validator_condition}
	\vspace{-1em}
\end{figure}

The thought process behind the validation mechanism for conditions is
illustrated in \autoref{fig:validator_condition}. First,
the validator extracts the condition from the \texttt{SELECT} clause and
interprets its meaning. Then, it analyzes the condition in the
\texttt{WHERE} clause. After interpreting both conditions, the validator
evaluates the execution response. In most cases, if the execution response
contains \texttt{None} or an empty set, it likely indicates an incorrect
condition (e.g., filtering incorrect values or using the wrong column in the
filter). The validator then identifies potential mistakes in the condition
and suggests ways to fix them. For example, it may recommend adding
conditions such as \texttt{"column A IS NOT NULL"} to filter out
\texttt{None} results or correcting mismatched conditions that were
misinterpreted by the LLM. Finally, the validator takes a decision on
whether the SQL query is considered correct or not.

\sstitle{Combined Validator}
Let \( v_s, v_c\) denote the instructions for the validator selection and
the validator condition, respectively. The validator agent, \( \pi_v \),
processes these instructions along with the input \( x \), the SQL query \(
s \), and its corresponding execution response~$er$, generating feedback
signals as follows:
\begin{equation}
    f_s = \pi_v(v_s, x, s, er)
\end{equation}
\begin{equation}
f_c = \pi_v(v_c, x, s, er)
\end{equation}
Here, \( f_s, f_c \) represent the feedback for the validator selection and
the validator condition, respectively.

\subsection{Fix Agent}
\label{sec:fix}
The Fix Agent refines SQL queries using the feedback obtained from the
Validator Agent. It processes the feedback to adjust and improve the SQL
query, aiming at ensuring that no further errors occur in the final output.

The Fix Agent takes the initial input $x$, the SQL query $s$ as generated by
the Planner, and a set of feedback signals as input to generate a corrected
SQL query. This process of the Fix Agent \(\pi_f\) is captured as:
\begin{equation}
    s_f = \pi_f\left(x, s, \left\{f \in \{f_s, f_c\} \mid f \text{ indicates
    an
    error}\right\}\right),
\end{equation}
where \(s_f\) is the corrected SQL query.

\subsection{Selection Agent}
\label{sec:selection}
The Selection Agent is an SLM that chooses the best SQL query from multiple
candidates based on the responses obtained when evaluating them over the
database.

The Selection Agent takes a prompt containing a list of SQL queries \( S =
\{s_1, s_2, \dots, s_K\} \) and their execution responses \( ER = \{er_1,
er_2, \dots, er_K\} \) as input, and outputs the index of the best query (or
indicates that no query is correct). When \(K > k\), the agent splits \( S
\) into smaller subsets (of size up to \( k \)), selects the best query in
each subset, and repeats this process until only one candidate remains.
Formally, the functionality of the Selection Agent \( \pi_s \) is captured
as:
\begin{equation}
s^* = \pi_s(S, ER),
\end{equation}
where \( s^* \) denotes the chosen query (and its index).

\subsection{Data Generation for Supervised Fine-Tuning}
\label{sec:data_creation_mechan}

To mitigate limitations in the reasoning with SLMs, we devise an approach to
generate training data for fine-tuning the models. To this end, we manually
label a small set of training examples per task, enabling SLMs to
adopt a structured reasoning process and reduce errors. For
each task, we construct training data based on the Spider and BIRD
datasets, as discussed in detail in our evaluation.
Since these datasets only provide input questions~$x$ and ground-truth SQL
queries~$\hat{s}$, we extend them with additional annotations: the planning
process~$p$ for the Planner, validator feedback $v_s, v_c$ for different
validation types, and corrected SQL queries based on the feedback. The data
creation  process follows three structured steps:

\smallskip
\textit{1) Manual Labeling:} We annotate up to five representative
examples for each task manually. These high-quality examples act as
references to guide each agent in performing small, incremental reasoning
steps, thereby minimizing error propagation.

\smallskip
\textit{2) Few-Shot Prompting:} Building on the manually labeled
examples, we use few-shot prompting techniques with OpenAI's GPT-4o-mini to
generate additional training samples for the rest of the dataset. This
approach ensures broader coverage of various scenarios in the dataset.

\smallskip
\textit{3) Fine-Tuning on SLMs:} We fine-tune SLMs using the
prompt-response pairs generated during the few-shot prompting phase. The
fine-tuning process focuses on optimizing the completion part of the
model's output. Here, we adopt a supervised fine-tuning loss, which is
computed as:
\begin{equation}
\label{eq:l_completion}
    \mathcal{L}_{\text{completion}} = -\sum_{t=C+1}^{\tau} \log P_{\theta}(y_t \mid y_{<t}, \chi),
\end{equation}
where $C$ is the token index marking the end of the prompt; $\tau$
represents the total number of tokens in the sequence which includes the
prompt and the completion tokens; $y_t$ denotes the target token at position
$t$ within the completion; $\chi$ represents the prompt tokens; and
$P_{\theta}$ is the probability distribution over the vocabulary predicted
by the model parameters $\theta$.

\section{Reinforcement Learning from Execution Feedback}
\label{sec:rlef}

\begin{figure*}[!h]
	\centering
	\includegraphics[width=\linewidth]{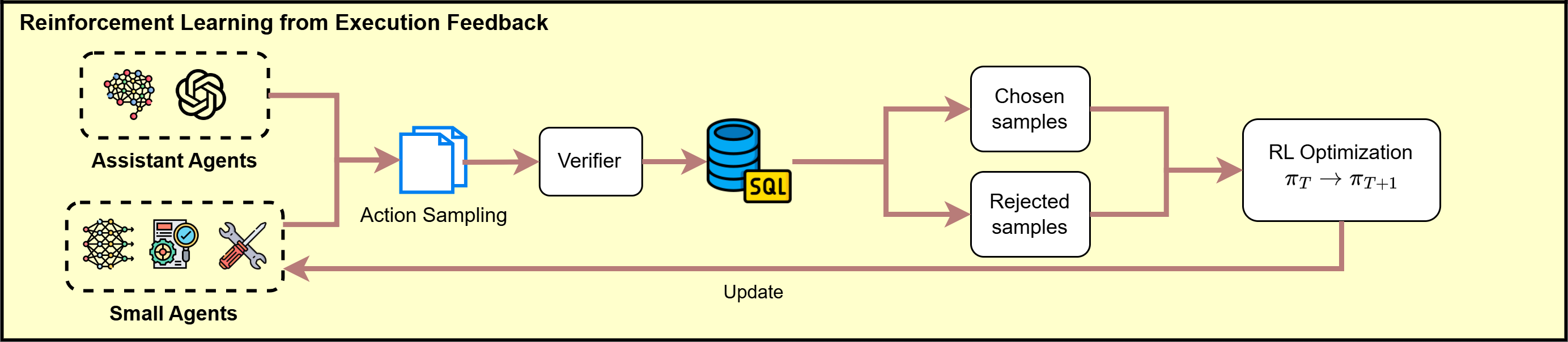}
	\caption{Overview of our approach to Reinforcement Learning from
	Execution Feedback.}
	\label{fig:framework_RLEF}
	\vspace{-.5em}
\end{figure*}

In the light of the limited capabilities of SLMs in terms of reasoning (R3)
and instruction following (R4), we propose an approach for reinforcement
learning from execution feedback (RLEF) to further refine the agents of the
\toolname framework (\autoref{fig:framework_RLEF}). RLEF targets cases where agents fail to generate correct SQL
queries, and relies on feedback from the evaluation of the queries to
identify issues and explore effective corrections.
Below, we first introduce
a respective model for the various agents (\autoref{sec:rl_modeling}),
before we target the identification of actions to train them
(\autoref{sec:rl_sampling}). Finally, we elaborate on the actual training
process (\autoref{sec:rl_training}).

\subsection{Modeling}
\label{sec:rl_modeling}

Given a user question and a database schema, we aim to generate an SQL query
that best fulfills the intent. We formulate the task as a
goal-augmented Partially Observable Markov Decision Process:
\[
M = (\mathcal{S}, A, T, R, G, O).
\]
The process definition includes the following components:
\begin{compactitem}
\item \( \mathcal{S} \) is a set of states;
\item \( A \subset V^L \) represents the action space sampled from the
language model's vocabulary \( V \) with $L$ as the maximum length of the
generated text;
\item \( T: \mathcal{S} \times A \rightarrow \mathcal{S} \) is the
transition function;
\item \( G \subset V^N \) denotes the goal space;
\item \( R: \mathcal{S} \times A \times G \rightarrow \mathbb{R} \)
represents the goal-conditioned reward function; and
\item \( O \) is a set of observations \( o \in O \).
\end{compactitem}
Note that the design of the reward function depends on
the specific algorithm used for training. For instance, the
DPO~\cite{rafailov2023direct} and ORPO\cite{hong-etal-2024-orpo} algorithms
do not require an explicit reward model, as the reward is inherently derived
from the language model itself~\cite{rafailov2023direct}. In this study, we
adopt the ORPO algorithm, so that, in the remainder, we use the
terms `chosen action' and `rejected action' to align with the algorithm's
framework.

We instantiate the process for the different types of agents in the \toolname
framework, as follows:

\medskip
\noindent
\textit{For the Planner}, the goal, observations, and actions are defined as:
\begin{compactitem}
    \item $G$ is the set of completions containing correct SQL queries.
    \item The observation $o_p \in O_p$ is the input sample $x$.
    \item The chosen action is a completion with a correct SQL query
    matching the ground-truth execution; the rejected action is a query
    resulting in a syntax error or differing execution.
\end{compactitem}

\medskip
\noindent
\textit{For the Validator}, the instantiation is done as follows:
\begin{compactitem}
    \item $G$ is the set of completions that enables the Fix Agent to modify
    the SQL query to a correct version.
    \item The observation $o_v \in O_V$ is the input prompt to the
    validator, consisting of $x$, the execution response of $x$, and an
    validation instruction $v_i \in \{v_s, v_c\}$.
    \item The chosen action is feedback that correctly identifies issues and
    assists the Fix Agent in modifying the SQL query. The rejected action is
    feedback that is either incorrect or insufficiently useful for the Fix
    Agent.
\end{compactitem}

\medskip
\noindent
\textit{For the Fix Agent}, we obtain the following realization:
\begin{compactitem}
    \item $G$ is the set of correct SQL queries.
    \item The observation $o_f \in O_F$ consists of the input sample $x$, an
    execution response, and a list of feedback signals indicating that the
    SQL is incorrect.
    \item The chosen action is an SQL query that matches
    the ground-truth result. The rejected action is a query that results in
    a syntax error or deviates from the ground truth.
\end{compactitem}

\subsection{Action Sampling}
\label{sec:rl_sampling}

The action space per agent is very large, with a size of $|V|^L$,
so that the identification of effective actions to train the agents is
important. However, random sampling or relying solely on the agent's
exploration of the action space will generally not yield effective actions.
We therefore adopt two strategies for action generation:
\begin{compactitem}
    \item Multinomial Decoding Strategy: We use a temperature parameter
    $\mathcal{T}$ to introduce controlled randomness during the decoding
    process of the text generated by SLMs. This helps the trained
    agent in exploring more effective actions.
    \item Advanced Agent Assistance: A larger, well-trained language model generates higher-quality actions for the task.
\end{compactitem}
Moreover, we introduce three assistant agents that leverage OpenAI
GPT-4o-mini during training:
\begin{compactitem}
    \item An Advanced Planner Agent often produces more accurate SQL queries
    compared to the smaller agent used in inference. For this agent, we rely
    on few-shot prompting.
    \item An Feedback Editor Agent modifies feedback to help the Fix Agent
    to correct SQL queries more effectively.
    \item An Advanced Fix Agent repairs SQL queries with a higher chance of
    success. It uses a prompt to refine the SQL query based on the provided
    feedback.
\end{compactitem}
For each agent, given an observation $o$, we generate a set of actions using
the multinomial decoding strategy and the associated assistant agent
$\mathcal{A}$. That is, we generate an action set $A_P \subset A$ of
size $K$, where $K-1$ actions are sampled from the multinomial decoding
strategy with a temperature $\mathcal{T}$, and one action is sampled from
the assistant agent $\mathcal{A}$.
The resulting action set $A_P$ may contain both, chosen actions
$\mathcal{C}$ and rejected actions $\mathcal{R}$. We then construct pairs
$\mathcal{P} = \{(o, c, r) \mid c \in \mathcal{C}, r \in
\mathcal{R}\}$. If $A_P$ contains only chosen actions ($\mathcal{C}$) or
only rejected actions ($\mathcal{R}$), the training sample is
ignored.

\subsection{Agent Alignment Training}
\label{sec:rl_training}
Using a set of chosen-rejected pairs from our sampling strategy, we improve
each agent with ORPO \cite{hong-etal-2024-orpo}. We choose ORPO for its
streamlined approach, which simplifies training and reduces computational
complexity compared to methods like PPO \cite{schulman2017proximal} or DPO
\cite{schulman2017proximal}. By integrating supervised fine-tuning loss with
an odds ratio-based penalty, ORPO preserves the accuracy gains from
supervised fine-tuning (SFT), while aligning model outputs to preferences.
This effectively avoids catastrophic forgetting as commonly seen, when
reinforcement learning follows SFT.

\begin{algorithm}[!h]
\renewcommand{\algorithmicrequire}{\textbf{Input:}}
\renewcommand{\algorithmicensure}{\textbf{Output:}}
\caption{Iterative Training with ORPO for Planner}
\label{alg:policy_planner}
\footnotesize
\begin{algorithmic}[1]
\Require \\
    Training dataset $\mathcal{D} = \{(x_i, \hat{s}_i) \mid i = 1, \dots, N\}$. \\
    Initial policy model $\pi_P$.\\
    Advanced Planner $\mathcal{A_P}$, Temperature $\mathcal{T}$, No. iterations $T$.
\Ensure
    Aligned policy for Planner $\pi_{p, T}$.

\State Initialize $\pi_{p, 0} \gets \pi_P$

\For{$i \gets 1$ \textbf{to} $T$}
    \State $D_p \gets \emptyset$

    \For{$(x, \hat{s}) \in \mathcal{D}$}
        \State $\textit{true\_response} \gets \textit{execute}(\hat{s})$
        \State $A \gets \textit{Sampling}(x, \pi_p, \mathcal{A_P}, \mathcal{T})$
        \State $\mathcal{C} \gets \emptyset$, $\mathcal{R} \gets \emptyset$ \Comment{Initialize chosen/rejected actions}

        \For{$a \in A$}
            \State $\textit{pred\_response} \gets \textit{execute}(a)$
            \If{$\textit{true\_response} = \textit{pred\_response}$}
                \State $\mathcal{C} \gets \mathcal{C} \cup \{a\}$
            \Else
            \State $\mathcal{R} \gets \mathcal{R} \cup \{a\}$
            \EndIf
        \EndFor

        \If{$|\mathcal{C}| > 0$}
    \If{$|\mathcal{R}| = 0$}
        \State $\mathcal{R} \gets \{""\}$ \Comment{Add empty string if $\mathcal{R}$ is empty}
    \EndIf
    \For{$c \in \mathcal{C}$}
        \For{$r \in \mathcal{R}$}
            \State $D_p \gets D_p \cup \{(x, c, r)\}$
        \EndFor
    \EndFor
\EndIf

    \EndFor

    \State Update policy $\pi_p$ using ORPO:
    $
    \pi_{p, i+1} \gets \text{ORPO}(\pi_p, D_p)
    $

\EndFor
\State \textbf{return} $\pi_{p,T+1}$
\end{algorithmic}
\end{algorithm}

The ORPO algorithm introduces a monolithic preference alignment approach
that eliminates the need for both, a reference model and a reward model,
thereby streamlining the preference optimization process. It employs a loss
function that integrates supervised fine-tuning loss with a log-odds ratio
penalty, enabling the model to effectively differentiate between chosen and
rejected actions. By default, ORPO computes SFT loss over the entire prompt
and completion. However, we modify the loss function to apply SFT loss only
to the completion, allowing the model to focus more directly on generating
high-quality outputs rather than learning the structure of the prompt.

\noindent
The objective function of ORPO is defined as:
\[
\mathcal{L}_{\text{ORPO}} = \mathbb{E}_{(x, y^w, y^l)} \left[ \mathcal{L}_{\text{completion}} + \lambda \cdot \mathcal{L}_{\text{OR}} \right]
\]
where:
\begin{compactitem}
\item  \( \mathcal{L}_{\text{completion}} \) is the standard supervised
fine-tuning loss that is computed only on the completion part, implemented
as
the negative log-likelihood of generating the favored response \(y^w\):
\[
\mathcal{L}_{\text{completion}} = - \frac{1}{m} \sum_{t=1}^{m} \log
P_\theta(y_t^w \mid x, y_{<t}^w)
\]

\item \( \mathcal{L}_{\text{OR}} \) is the log odds ratio penalty,
encouraging
the model to increase the odds of generating favored responses over
disfavored ones:
\[
\mathcal{L}_{\text{OR}} = - \log \sigma \left( \log
\frac{\text{odds}_\theta(y^w \mid x)}{\text{odds}_\theta(y^l \mid x)} \right)
\]

\item \(\text{odds}_\theta(y \mid x)\) is the likelihood odds of generating
the sequence \(y\), which is computed as:
\[
\text{odds}_\theta(y \mid x) = \frac{P_\theta(y \mid x)}{1 - P_\theta(y \mid
x)}
\]

\item \(\lambda\) is a hyperparameter controlling the weight of the log odds
ratio penalty.
\end{compactitem}

Using this mechanism, we iteratively enhance each agent starting with a
given policy \(\pi\). At each training stage, the policy \(\pi_t\) is
refined, which yields a policy \(\pi_{t+1}\). This process continues in a
loop until the
training dataset size stabilizes, which indicates that
further fine-tuning fails
to correct errors. These persistent errors are often
associated with
semantic issues, which are intrinsically linked to the limitations of small
language models.

The detailed training procedure for the Planner Agent is presented in
\autoref{alg:policy_planner}. The training process for the Fix Agent is
similar, except that the observation $x$ includes the generated SQL query
and feedback signals from the Validator Agent.

\begin{algorithm}[!h]
\renewcommand{\algorithmicrequire}{\textbf{Input:}}
\renewcommand{\algorithmicensure}{\textbf{Output:}}
\caption{Action Selection Process for Validator Agent}
\label{alg:validator_training}
\footnotesize
\begin{algorithmic}[1]
\Require \\
    Observation $o$; Action sets $A_s$, $A_c$ sampled with observation $o$.\\
    Ground-truth SQL $\hat{s}$; Planner SQL response $s$.\\
    Fix agent $\pi_{F, t-1}$.
\Ensure
    Chosen sets: $\mathcal{C}_s$, $\mathcal{C}_c$.
    Rejected sets: $\mathcal{R}_s$, $\mathcal{R}_c$.

\State $\textit{true\_response} \gets \textit{execute}(\hat{s})$
\State $\textit{planner\_response} \gets \textit{execute}(s)$

\If{$\textit{planner\_response} = \textit{true\_response}$}
\For{$a \in \{A_s, A_c\}$}
    \If{$a$ indicates the SQL is correct.}
        \State Add $a$ to the corresponding chosen set $\mathcal{C}$
    \Else
        \State Add $a$ to the corresponding rejected set $\mathcal{R}$
    \EndIf
\EndFor

\EndIf

\For{$(a_s, a_c) \in (A_s, A_c)$}
    \State $s_f \gets \pi_{F, t-1}(o, s, \{a \mid a \in \{a_s, a_c\},$
    \State \hspace{2em} $a \text{ indicates an error}\})$

    \State $\textit{pred\_response} \gets \textit{execute}(s_f)$
    \If{$\textit{pred\_response} = \textit{true\_response}$}
        \State Add $a_s$ to $\mathcal{C}_s$, $a_c$ to $\mathcal{C}_c$
    \Else
        \State Add $a_s$ to $\mathcal{R}_s$, $a_c$ to $\mathcal{R}_c$
    \EndIf
\EndFor
\State \textbf{return} $\mathcal{C}_s, \mathcal{C}_c,\; \mathcal{R}_s, \mathcal{R}_c$

\end{algorithmic}
\end{algorithm}

The training process for the Validator Agent differs as it does not generate
SQL queries to compare with the ground-truth query. To determine the chosen
and rejected actions for the Validator Agent $\pi_{V, t}$, we incorporate
the Fix
Agent from the previous aligned step, $\pi_{F, t-1}$. If the actions
generated by the Validator Agent help the Fix Agent write a correct SQL
query, the
actions are chosen; otherwise, they are rejected. This process is described
in \autoref{alg:validator_training}.

Note that we do not handle cases where
the  $\textit{planner\_response} \neq \textit{true\_response}$, as
determining the incorrect feedback among the two types is challenging. For
instance, even with an incorrect SQL query, the selection feedback might be
correct, while the error lies solely in the condition feedback.

\section{Empirical Evaluation}
\label{sec:exp}

In this section, we conduct experiments with the aim of answering the
following research questions:
\begin{compactitem}
\item[(RQ1)] Does \toolname outperform baseline methods in terms of accuracy and
efficiency?
\item[(RQ2)] How robust is \toolname in handling noisy and realistic queries?
\item[(RQ3)] Do hyper-parameters strongly affect our model?
\item[(RQ4)] How does RLEF improve SQL generation accuracy?
\item[(RQ5)] How do different RL algorithms impact accuracy?
\item[(RQ6)] Is our thought process for small agents important?
\item[(RQ7)] How does \toolname perform in different domains?
\item[(RQ8)] How do SQL characteristics influence \toolname?

\end{compactitem}

\subsection{Experimental Setup}

\sstitle{Database} We rely on two English Text2SQL benchmarks:
Spider~\cite{yu-etal-2018-spider} and BIRD~\cite{li2024can}.

\begin{compactitem}
\item 
\textit{Spider} is a popular benchmark for NL2SQL translation, consisting
of 200 databases with multiple tables that cover 138 diverse domains. Spider
contains 7000 samples in a training set, a development set with 1024
samples, and a hidden test set.

\item
\textit{Spider Variants:} Spider-DK~\cite{gan-etal-2021-exploring},
Spider-Syn~\cite{gan2021towards}, as well as
Spider-Realistic~\cite{deng-etal-2021-structure} add real-world challenges
to the original Spider dataset by introducing synonyms, abbreviations, and
modified column/value names. Dr.Spider~\cite{chang2023drspider} adds 17
perturbations across databases, questions, and SQL queries
(schema/value-synonyms, sort-order, numerical changes), providing a more
holistic robustness test.

\item
\textit{BIRD} contains 95 databases, cumulatively accounting for 33.4GB
across 37 professional domains. BIRD contains 9428 samples in
training set, a development set with 1534 samples and a hidden test set.
BIRD is more challenging, with each of BIRD's databases containing around
549K rows on average, compare to Spider's limited capacity of just 2k
rows. Also, BIRD offers evidence for a specific sample to facilitate
the generation of the right SQL query.
\end{compactitem}

\sstitle{Evaluation Metrics} For the Spider benchmark, we measure the
execution accuracy (EX). It evaluates whether the predicted and ground-truth
SQL queries yield the same execution results on the database. However, EX
can occasionally produce false positives, when incorrect SQL queries
accidentally yield the same results as the ground truth. To address this,
the Spider benchmark also utilizes the Test-Suite accuracy
(TS)~\cite{li2024codes}. It assesses whether the predicted SQL query passes
execution across multiple database instances generated through automated
augmentations, making it a more reliable metric by reducing false positives.

The BIRD benchmark primarily relies on execution accuracy (EX) as its
evaluation metric. Additionally, BIRD introduces the Valid Efficiency Score
(VES) to evaluate the execution efficiency of correctly generated SQL
queries. Unlike EX, in VES, the score for a correct SQL query is calculated
as the execution time of the ground-truth divided by the execution time
of the predicted SQL. If the predicted SQL is more efficient, its VES score
surpasses~EX.

\sstitle{Baselines} We compare our method with the following baselines:

\begin{compactitem}
\item
\textit{Open-Source Fine-Tuning Models.} T5~\cite{li2024can} is a
    fine-tuned baseline for Text2SQL tasks, tested at various scales (Base,
    Large, 3B). CodeS~\cite{li2024codes} is an open-source model (1B--15B
    parameters) incrementally pre-trained on NL2SQL data to specialize in
    Text2SQL. We do not compare with DeepSeek-R1, even though it is
    considered an open-source model and has a large parameter size
    (671B), because its API was temporarily unavailable at the time of our
    research due to excessive traffic.

\item
\textit{Closed-Source API-Based Methods.} ChatGPT + CoT~\cite{li2024can}
enhances SQL generation with chain-of-thought reasoning for improved
multi-step problem-solving. DIN-SQL~\cite{pourreza2023din} adopts a
decomposed in-context learning approach, breaking queries into subtasks such
as schema linking, query classification, and self-correction.
DAIL-SQL~\cite{gao2023text} integrates prompt engineering with systematic
fine-tuning, optimizing example selection based on question-query similarity
and a balanced organizational strategy. MAC-SQL~\cite{wang2023mac} employs a
multi-agent framework featuring a Decomposer for sub-question processing.
Finally, CHESS~\cite{talaei2024chess} utilizes a modular pipeline that
includes entity/context retrieval, schema selection, and SQL generation,
leveraging LLMs for schema pruning to minimize noise.

\end{compactitem}

\sstitle{Setup}
The Schema Insight Agent uses
RoBERTa-large~\cite{liu2019roberta} (355M parameters), while the Planner and
Selection Agents are fine-tuned on LLaMA-3.2 3B. The Validator and Fix
Agents are fine-tuned on LLaMA-3.2 1B, resulting in a total parameter size
of \toolname of {9B}. All agents, except the Schema Insight Agent, utilize
bfloat16 precision and FlashAttention~\cite{dao2022flashattention}.

For inference, all experiments were conducted on a machine equipped with an
A5000 GPU featuring 24GB of memory. The models were deployed using the VLLM
framework~\cite{kwon2023efficient}. For supervised fine-tuning, we set the
learning rate to $2.0 \times 10^{-5}$, a batch size of 128, and train the
model for 4 epochs. For ORPO training, we set the learning rate to $5 \times
10^{-6}$, $\lambda$ to 0.5, and a batch size of 64, and the number of epochs
is 1 or within 800 steps. For the Planner agent, we sample $K = 10$
solutions, one by greedy decoding and the others from multinomial sampling
with a temperature $\mathcal{T} = 1.0$.

\subsection{End-to-end comparison}
\autoref{table:end2end_bird_dev} and \autoref{table:end2end_spider_dev}
present the performance of \toolname in comparison to other baselines. In these
tables, CHESS$_{open-source}$  refers to the CHESS method that uses
Llama-3-70B and Fine-tuned DeepSeek in its workflow.

\sstitle{Accuracy}
\toolname shows a strong performance on both, the Spider and BIRD datasets. On
the Spider development set, \toolname reaches an EX\% of 87.1, outperforming
other methods, including larger models such as CodeS-15B and MAC-SQL +
GPT-4. On the BIRD development dataset, the \toolname model achieves an EX\% of
64.73 and a VES\% of 66.75. This performance is comparable to CHESS, which
incorporates multiple modules leveraging OpenAI's GPT-4-turbo. Overall, \toolname
outperforms all open-source models in both EX\% and VES\%, while being
on-par with closed-source methods that rely heavily on proprietary models,
such as OpenAI GPT.

\sstitle{Time}
We compare inference time only with other open-source fine-tuning methods,
as closed-source approaches host LLMs on unknown physical devices.
We note that the inference time of \toolname is slower than that of CodeS-15B due
to (1) structural differences between Llama and CodeS, and (2) the presence
of multiple agents in \toolname. Additionally, the inference time of \toolname on
Spider and BIRD differs significantly, averaging 13.9 s/sample on Spider
compared to 22.03 s/sample on BIRD. This is because BIRD contains many large
databases, leading to a higher execution times.

\begin{table}[t]
\caption{Evaluation of \toolname on BIRD dev.}
\label{table:end2end_bird_dev}
\vspace{-1em}
\centering
\resizebox{0.95\columnwidth}{!}{
\begin{tabular}{|l|l|l|l|l|}
\hline
\textbf{Methods}               & \textbf{\#Params} & \textbf{EX\%} & \textbf{VES\%} & \textbf{Avg. Time} \\
                               &                       &               &                & \textbf{(s/sample)}          \\ \hline
\multicolumn{5}{|c|}{\textbf{Closed-source API-based methods}}                                         \\ \hline
ChatGPT + CoT & - &  36.64 & 42.30 & \\ \hline
DIN-SQL + GPT-4                & -                     & 50.72         & 58.79          & -                             \\ \hline
DAIL-SQL + GPT-4               & -                     & 54.76         & 56.08          & -                             \\ \hline
MAC-SQL + GPT-4                          & -                     & 59.59         & 66.39         & -                             \\ \hline
GPT-4o + CoT                          & -                     & 54.43         &  58.73        & -                             \\ \hline
CHESS + GPT-4                          & -                     & \textbf{65.00}         & \textbf{66.69}         & -                             \\ \hline

\multicolumn{5}{|c|}{\textbf{Open-source fine-tuning methods}}                                       \\ \hline
Fine-tuned T5-3B               & 3B                    & 23.34         & 25.57          & -                             \\ \hline
SFT Llama2-7B                  & 7B                    & 45.37         & 46.98          & -                             \\ \hline
SFT Llama2-13B                 & 13B                   & 53.91         & 58.77          & -                             \\ \hline
CodeS-1B         & 1B                    & 50.46         & 51.07          & 0.69                         \\ \hline
CodeS-3B        & 3B                    & 55.02         & 56.54          & 1.06                         \\ \hline
CodeS-7B        & 7B                    & 57.17         & 58.80          & 1.87                         \\ \hline
CodeS-15B        & 15B                   & 58.47         & 59.87          & 3.52                         \\ \hline
CHESS$_{open-source}$       & 33B + 70B                   & 59.86         & -          & -                         \\ \hline
\multicolumn{5}{|c|}{\textbf{Ours}}                                                     \\ \hline
\toolname                       & 9B                & \textbf{64.73}         & \textbf{66.75}          & 22.03                          \\ \hline
\end{tabular}}
\vspace{-0em}
\end{table}

\begin{table}[t]
\caption{Evaluation of \toolname on Spider dev.}
\label{table:end2end_spider_dev}
\vspace{-1em}
\centering
\resizebox{0.95\columnwidth}{!}{
\begin{tabular}{|l|l|l|l|l|}
\hline
\textbf{Methods}               & \textbf{\#Params} & \textbf{EX\%} & \textbf{TS\%} & \textbf{Avg. Time} \\
                               &                       &               &                & \textbf{(s/sample)}          \\ \hline

\multicolumn{5}{|c|}{\textbf{Closed-source API-based methods}}                                         \\ \hline
GPT-4 (few-shot)               & -                 & 76.8          & 67.4          & -                             \\ \hline
C3 + ChatGPT                   & -                 & 81.8          & 71.4          & -                             \\ \hline
DIN-SQL + GPT-4                & -                 & 82.8          & 74.2          & -                             \\ \hline
DAIL-SQL + GPT-4               & -                 & 83.1          & 76.6          & -                             \\ \hline
MAC-SQL + GPT-4                & -                 & \textbf{86.75}         & -             & -                             \\ \hline
\multicolumn{5}{|c|}{\textbf{Open-source fine-tuning methods}}                                       \\ \hline
T5-3B + PICARD                 & 3B                & 79.3          & 69.4          & -                             \\ \hline
RESDSQL-3B            & 3B                & 84.1          & 73.5          & -                             \\ \hline
SFT Llama2-7B                  & 7B                & 77.8          & 73.5          & -                             \\ \hline
SFT Llama2-13B                 & 13B               & 81.6          & 76.6          & -                             \\ \hline
CodeS-1B                       & 1B                & 77.9          & 72.2          & 0.45                          \\ \hline
CodeS-3B                       & 3B                & 83.4          & 78.1          & 0.71                          \\ \hline
CodeS-7B                       & 7B                & 85.4          & 80.3          & 1.22                          \\ \hline
CodeS-15B                      & 15B               & 84.9          & 79.4          & 2.38                          \\ \hline
\multicolumn{5}{|c|}{\textbf{Ours}}                                                     \\ \hline
\toolname                       & 9B            &  \textbf{87.1}         &  \textbf{82.3}        &      13.9                     \\
\hline
\end{tabular}}
\vspace{-.5em}
\end{table}

\begin{table}[htbp]
	\caption{GPU memory requirement for serving LMs.}
	\label{tab:memory}
	\vspace{-1em}
	\centering
	\resizebox{0.95\columnwidth}{!}{
		\begin{tabular}{lcc}
			\toprule
			\textbf{Method} & \textbf{Float precision} & \textbf{GPU mem.
				(Gb)} \\
			\midrule
			CodeS-1B  & float16  & 2.4  \\
			CodeS-3B & float16  & 7.2  \\
			CodeS-7B  & float16 & 16.8 \\
			CodeS-15B  & float16 & 36   \\
			RESDSQL-3B\cite{li2024dawn} & float32 & 24.66 \\
			RESDSQL-3B + NatSQL\cite{li2024dawn} & float32 & 21.59 \\
			\toolname Planner & bfloat16 & 7.6 \\
			\toolname & bfloat16 & 21.2 \\
			\bottomrule
	\end{tabular}}
	\vspace{-0em}
\end{table}

\begin{table}[t]
\caption{Evaluation \toolname on Spider variants.}
\label{tab:spider_variants}
\vspace{-1em}
\centering
\resizebox{0.95\columnwidth}{!}{
\begin{tabular}{lccccc}
\toprule
\textbf{Model} & \multicolumn{2}{c}{\textbf{Spider-Syn}} & \multicolumn{2}{c}{\textbf{Spider-Realistic}} & \textbf{Spider-DK} \\
 & \textbf{EX\%} & \textbf{TS\%} & \textbf{EX\%} & \textbf{TS\%} & \textbf{EX\%} \\
\midrule
T5-3B + PICARD & 69.8 & 61.8 & 71.4 & 61.7 & 62.5 \\
REDSQL-3B\cite{li2023resdsql} & 76.9 & 66.8 & 81.9 & 70.1 & 66 \\
ChatGPT & 58.6 & 48.5 & 63.4 & 49.2 & 62.6 \\
SQL-PaLM & 70.9 & 66.4 & 77.4 & 73.2 & 67.5 \\
CodeS-1B & 66.5 & 59.3 & 70.9 & 61.8 & 64.7 \\
CodeS-3B & 75.7 & 69 & 79.9 & 74.4 & 71.8 \\
CodeS-7B & 76.9 & 70 & 82.9 & 77.2 & 72 \\
CodeS-15B & 77 & 69.4 & 83.1 & 75.6 & 70.7 \\
\midrule
\toolname   & \textbf{78.5} & \textbf{72.0} & \textbf{84.4} & \textbf{78.0} & \textbf{74.0} \\
\bottomrule
\end{tabular}}
\vspace{-0em}
\end{table}

\sstitle{Model Complexity}
The autoregressive models are served using VLLM, with memory requirements
determined by model size and floating-point precision, as shown in
\autoref{tab:memory}. The \toolname Planner requires 7.6 GB of GPU memory,
comparable to CodeS-3B, while the full \toolname framework requires 21.2 GB, so
that, unlike CodeS-15B, it can be deployed on a single GPU. While the
\toolname framework has higher memory demands than most of the other open-source
methods, it delivers superior execution accuracy and inference efficiency,
as shown in \autoref{table:end2end_bird_dev}
and \autoref{table:end2end_spider_dev}.

\begin{table*}[t]
	\caption{Evaluation of \toolname on Dr.Spider.}
	\label{tab:dr_spider}
	\vspace{-1em}
	\centering
	\resizebox{0.8\linewidth}{!}{
		\begin{tabular}{|l|l|r|r|r|r|r|r|}
			\hline
			\textbf{Type} & \textbf{Perturbation} & \textbf{\#Sam.} &
			\textbf{REDSQL-3B} & \textbf{ChatGPT} & \textbf{CodeS-7B} &
			\textbf{CodeS-15B} & \textbf{\toolname} \\
			&  & & \textbf{\cite{li2023resdsql}} &
			\textbf{+ZeroNL2SQL\cite{gu2023interleaving}} &  &  &   \\
			\hline
			\multirow{4}{*}{DB} & schema-synonym & 2619 & 68.3 & 69.8 & 67.2
			& 66.9 & 73.0 \\
			& schema-abbreviation & 2853 & 70.0 & 74.8 & 76.8 & 78.7 & 79.2
			\\
			& DBcontent-equivalence & 382 & 40.1 & 56.8 & 46.9 & 47.6 & 52.1
			\\
			& \textbf{Average} & - & 59.4 & 67.1 & 63.6 & 64.4 &
			\textbf{74.7} \\ \hline
			\multirow{8}{*}{NLQ} & keyword-synonym & 953 & 72.4 & 74.0 &
			73.0 & 73.5 & 73.1 \\
			& keyword-carrier & 399 & 83.5 & 88.2 & 91.5 & 91.7 & 89.5 \\
			& column-synonym & 563 & 63.1 & 62.7 & 63.2 & 64.7 & 63.8 \\
			& column-carrier & 579 & 63.9 & 71.7 & 80.7 & 79.1 & 75.3 \\
			& column-attribute & 119 & 71.4 & 70.6 & 63.0 & 68.9 & 72.3 \\
			& column-value & 304 & 76.6 & 76.0 & 73.7 & 76.3 & 74.0 \\
			& value-synonym & 506 & 53.2 & 70.6 & 72.7 & 71.9 & 71.3 \\
			& multitype & 1351 & 60.7 & 66.4 & 69.5 & 69.4 & 68.1 \\
			& others & 2819 & 79.0 & 79.4 & 81.5 & 81.2 & 81.2 \\
			& \textbf{Average} & - & 69.3 & 73.2 & 74.3 & \textbf{76.3} &
			75.5 \\ \hline
			\multirow{5}{*}{SQL} & comparison & 178 & 82.0 & 73.6 & 77.5 &
			71.9 & 77.5 \\
			& sort-order & 192 & 85.4 & 80.2 & 81.8 & 84.9 & 81.2 \\
			& nonDB-number & 131 & 85.5 & 92.4 & 90.1 & 84.0 & 90.8 \\
			& DB-text & 911 & 74.3 & 80.7 & 80.5 & 80.7 & 79.3 \\
			& DB-number & 410 & 88.8 & 86.1 & 84.9 & 85.9 & 91.2 \\
			& \textbf{Average} & - & 83.2 & 82.6 & 83.0 & 81.5 &
			\textbf{87.0} \\ \hline
			\textbf{All} & Global average & - & 71.7 & 74.9 & 75.0 & 75.1 &
			\textbf{76.0} \\ \hline
	\end{tabular}}
	\vspace{-.5em}
\end{table*}

\subsection{Evaluation on Robustness Benchmarks}

In this experiment, we evaluate the execution accuracy of \toolname on noisy and
more realistic questions using Spider variants and the Dr.Spider datasets.
\autoref{tab:spider_variants} showcases the performance of \toolname on
Spider-Syn, Spider-Realistic, and Spider-DK, highlighting its capability to
handle diverse and challenging query variations.

To further assess \toolname's
robustness, we evaluate it on Dr.Spider. Recall that this benchmark
comprising 17 types of perturbations across questions,
database schemas, and SQL queries, including schema-synonym replacements,
value synonyms, and abbreviation updates.
However, our model is trained exclusively on the original Spider dataset,
i.e., without any question or database augmentation.

\autoref{tab:dr_spider} compares the execution accuracy of \toolname with other
baseline approaches. \toolname outperforms all baselines on DB and SQL
perturbations. However, on NLQ perturbations, \toolname achieves a slightly lower
accuracy (75.5\%) compared to CodeS-15B (76.3\%). This highlights the
limitations of SLMs in understanding natural language, particularly when
handling synonyms and linguistic variations. Overall, \toolname proves robust and
adaptable to noisy, realistic queries, making it effective for real-world
Text2SQL tasks.

\subsection{Hyperparameter Sensitivity}

\sstitle{Number of candidates}
As the Planner produces multiple responses, increasing the number of
candidates may improve the chances of finding the correct SQL query. We explore the selection effectiveness as follows. We evaluate the
upper bound by measuring recall when generating \(
K \) SQL queries and assessing the Selection Agent's accuracy in identifying
the correct one.

\begin{figure*}[!ht]
    \centering
    \begin{subfigure}{0.3\linewidth}
        \centering
        \includegraphics[width=0.98\linewidth]{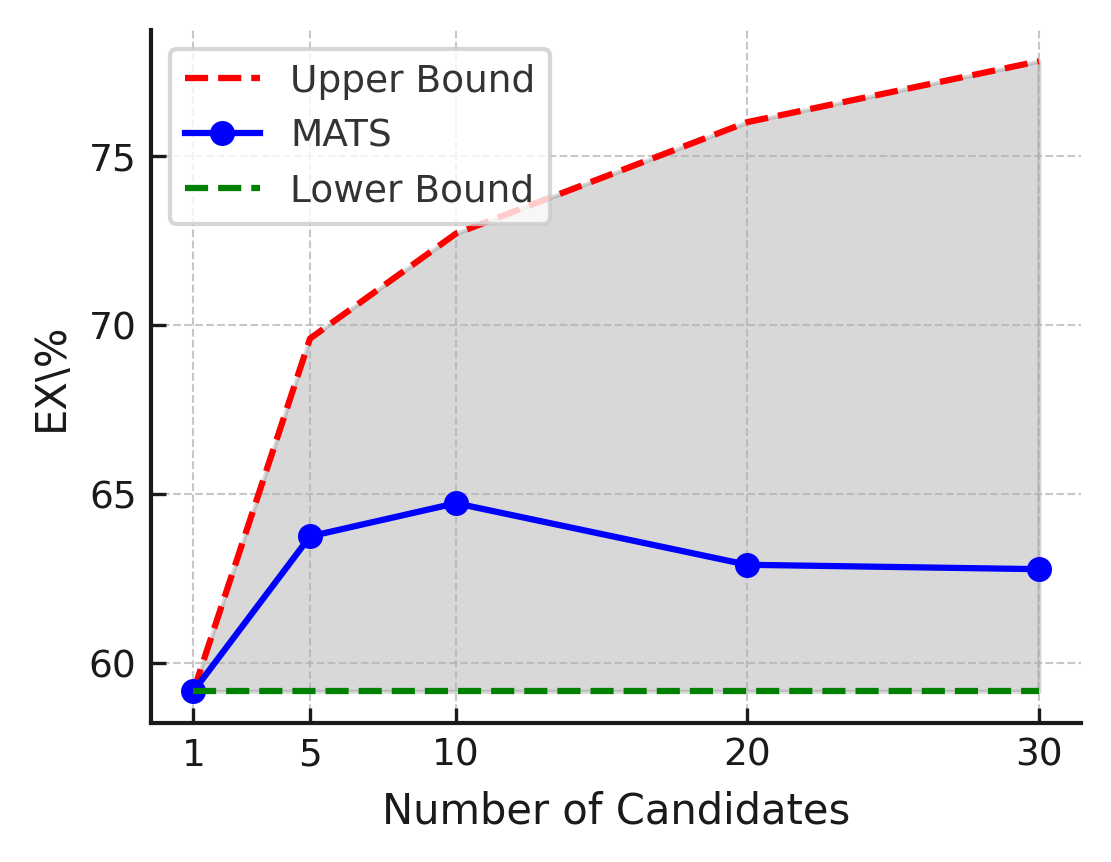}
        \vspace{-0.5em}
        \caption{\toolname EX\% on BIRD Dev with increasing number of candidates.}
        \label{fig:selection_effect}
    \end{subfigure}
    \hfill
    \begin{subfigure}{0.3\linewidth}
        \centering
        \includegraphics[width=0.98\linewidth]{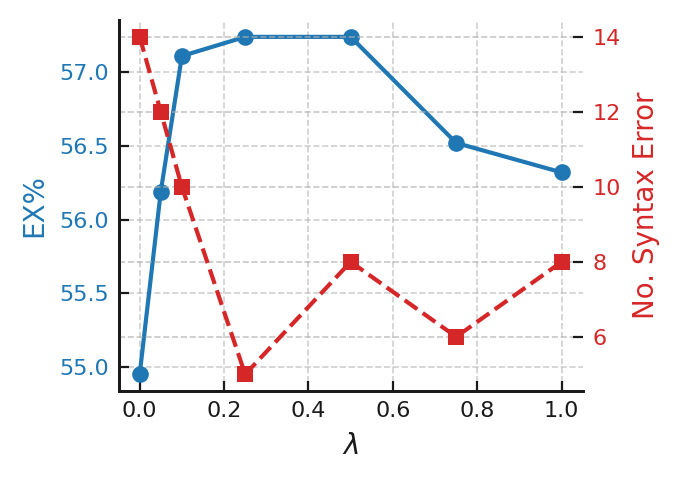}
        \vspace{-0.5em}
        \caption{EX\% on BIRD Dev after finetuning the Planner with RLEF
        Iteration 1 while varying $\lambda$.}
        \label{fig:lambda_sensitivity}
    \end{subfigure}
    \hfill
    \begin{subfigure}{0.3\linewidth}
        \centering
        \includegraphics[width=0.98\linewidth]{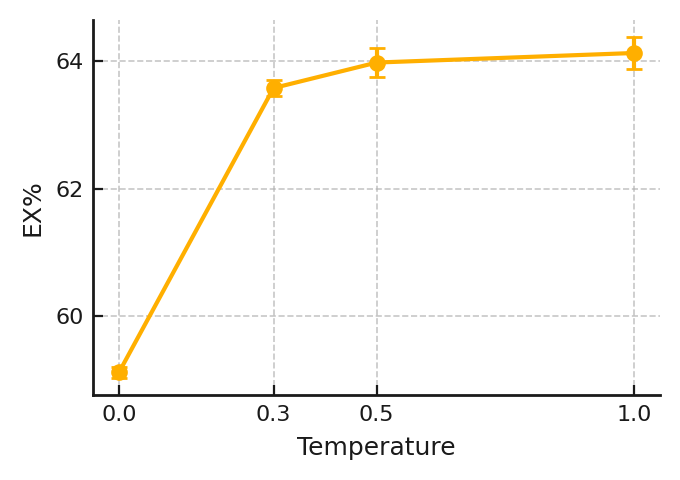}
        \vspace{-0.5em}
        \caption{EX\% on BIRD Dev with different temperatures for candidate
        generation.}
        \label{fig:temperature_sensitivity}
    \end{subfigure}
    \vspace{-.5em}
    \caption{Comparison of different factors affecting EX\% on BIRD Dev.}
    \label{fig:combined_figure}
     \vspace{-.5em}
\end{figure*}

\autoref{fig:selection_effect} presents the execution
accuracy of \toolname on the BIRD development dataset as the number of candidates
increases. The upper bound represents the recall, indicating the maximum
possible accuracy if the correct query is always included in the candidate
set. Specifically, if at least one of the \( K \) generated candidates
contains the correct SQL query, the sample is considered correct. As
expected, increasing \( K \) improves recall, leading to a higher upper
bound. However, the accuracy of \toolname does not consistently follow the same
trend. While recall reaches 77.8\% at \( K = 30 \), \toolname achieves only
62.78\%, highlighting a gap between potential and actual execution
accuracy. The reason is likely that more incorrect SQL queries are
generated, increasing the
likelihood that the Selection Agent chooses an incorrect query.

\sstitle{$\lambda$ to balance the SFT loss and the log odds ratio penalty}
\autoref{fig:lambda_sensitivity} presents the impact of varying $\lambda$ on
EX\% and the number of syntax errors after finetuning the Planner with RLEF
Iteration 1. In this figure, $\lambda=0$ means there is no log odds ratio
penalty and only SFT is applied. Introducing $\lambda > 0$ with the log odds
ratio greatly improves accuracy, reaching a peak at $\lambda=0.25$.
Meanwhile, the number of syntax errors decreases significantly for $\lambda
> 0.2$. However, our results also highlight that the choice of $\lambda$
needs to be made carefully to
balance accuracy and syntactical correctness.

\sstitle{Temperature for candidate generation}
\autoref{fig:temperature_sensitivity} reports the mean and standard
deviation (std) of EX\% across different temperature settings during
candidate generation. Lower temperatures exhibit lower std, indicating more
stable outputs. However, they also make it harder to find a correct
solution, as the generated candidates are less diverse. In contrast, higher
temperatures increase exploration, making it easier to generate correct
solutions among the candidates, though at the cost of greater variability.
Note that even with temperature 0.0, the results remain slightly
non-deterministic due to underlying sampling mechanisms.

\subsection{Improvement after each RLEF Iteration}
\label{sec:alignment_vs_sft}

\begin{figure}[!ht]
	\centering
	\includegraphics[width=0.9\linewidth]{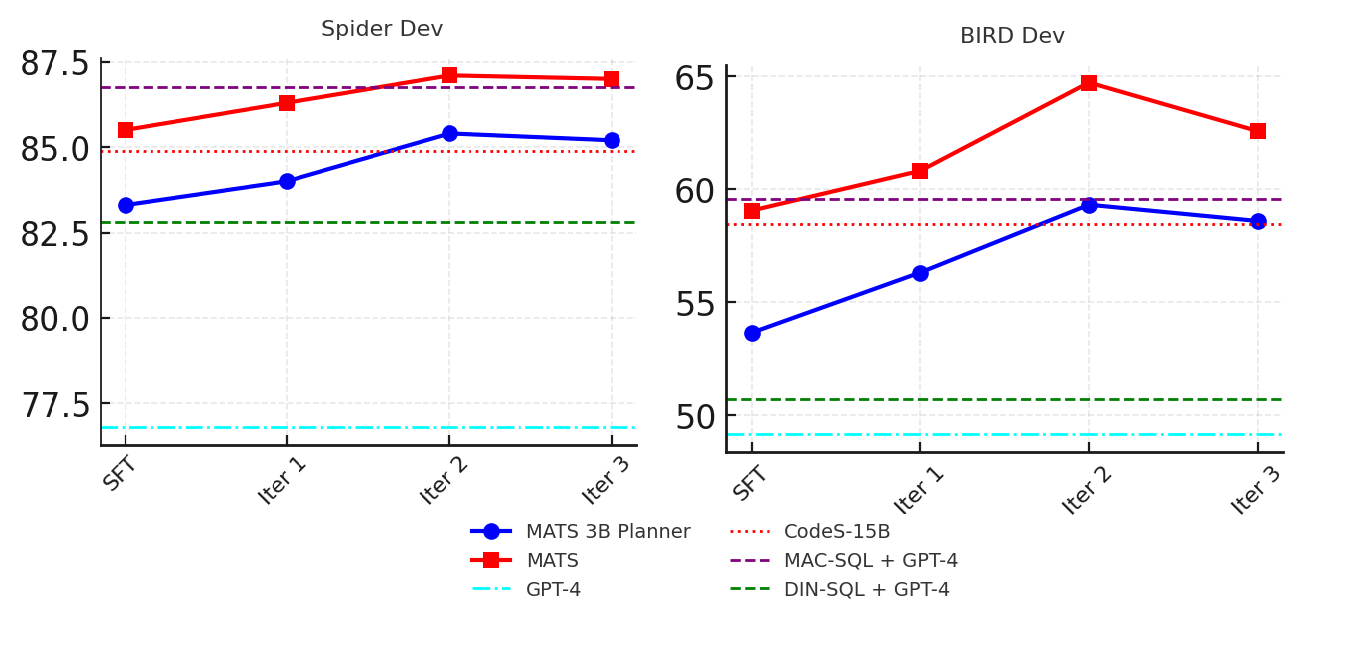}
	\vspace{-1em}
	\caption{Execution accuracy of \toolname Planner with greedy decoding per
	iteration of RLEF on BIRD Dev and Spider Dev. }
	\label{fig:rlef_iterations}
	\vspace{-1em}
\end{figure}

\autoref{fig:rlef_iterations} illustrates the impact of each RLEF iteration on the execution accuracy of both the Planner Agent and the entire framework. For the Planner Agent, we report accuracy using greedy decoding. The results demonstrate that RLEF significantly enhances the ability to generate correct SQL queries. On the BIRD dev dataset, RLEF improves the Planner Agent's EX\% from 53.65 to 59.32, while on the Spider dev dataset, it increases from 83.3 to 85.4. The entire framework, \toolname, also benefits from RLEF, exhibiting notable improvements across iterations. On Spider dev, \toolname progresses from an initial EX\% of 85.5 to 87.1, while on BIRD dev, it improves from 59.06 to a peak of 64.73 before slightly stabilizing. These findings confirm that execution feedback is both helpful and essential for improving the execution accuracy of small language models.

\subsection{Comparison of Alignment Algorithms}
\label{sec:compare_alignment}

We evaluate the effectiveness of different algorithms including
PPO, DPO, and ORPO in aligning the Planner policy. Here,
ORPO$_{original}$ is the original version of ORPO, and ORPO$_{ours}$ is our
modified version of ORPO with the loss computing on the completion part only
(see \autoref{sec:rlef}). For PPO, we use binary rewards, assigning 1 for
correct execution and 0 otherwise. For DPO, we set $\beta = 0.1$ with a
learning rate of \(5 \times 10^{-6}\). All models are trained for one epoch.

As shown in \autoref{tab:training_comparison}, our modified ORPO achieves
the highest execution accuracy (56.32\%), outperforming PPO and DPO.
Additionally, our modification for ORPO shows that ORPO$_{ours}$ consumes
less GPU Memory, while achieving better execution accuracy. PPO provides a
slight improvement over the SFT baseline,
increasing accuracy from 53.65\% to 54.95\%, but incurs a significantly high
training time (12.2 hours per epoch). DPO, in contrast, suffers from
catastrophic forgetting, causing accuracy to drop to 44.98\%. Even if this
issue is mitigated, its GPU memory consumption (67 GB) and training time
(1.8 hours) remain higher than ORPO.

Notably, ORPO eliminates the reference
model, reducing GPU memory usage by nearly half. This enables larger batch
sizes, resulting in faster training (1.5 hours per epoch) while maintaining
compatibility with limited GPU resources, making it a more practical choice
in constrained environments.

\begin{table}[!h]
\vspace{-.5em}
\centering
\caption{EX of different alignment training approaches on the \toolname planner
model after Iteration 1}
\label{tab:training_comparison}
\resizebox{0.95\columnwidth}{!}{
\begin{tabular}{|l|c|c|c|}
\hline
\textbf{Model} & \textbf{EX\%} & \textbf{GPU Memory} & \textbf{Training Time} \\
 &  & \textbf{(Gb)} & \textbf{(h/epoch)} \\ \hline
\toolname Planner (SFT only) & 53.65 & -&  - \\
\hline
+ PPO & 54.95 & \textbf{46} & 12.2 \\ \hline
+ DPO & 44.98 & 67 & 1.8 \\ \hline
+ ORPO$_{original}$ & 55.02 & 56 & 1.6 \\ \hline
+ ORPO$_{ours}$ & \textbf{56.32} & 52 & \textbf{1.5} \\ \hline
\end{tabular}}
\vspace{-0em}
\end{table}

\subsection{Planner Prompt Comparison}
\label{sec:planner_prompt_comparison}

In this experiment, we compare different prompting strategies for supervised
fine-tuning the Planner. The input for supervised fine-tuning includes both
the prompt and its corresponding completion. We construct three supervised
fine-tuning datasets with distinct prompting strategies: (1) No Thought,
where the ground-truth SQL queries from the BIRD Train dataset are directly
used as completions; (2) Chain-of-Thoughts, where SQL queries are generated
using CoT prompting on GPT-4o-mini and used for fine-tuning; and (3)
Few-shot Thoughts, where our proposed data generation mechanism, as
described in \autoref{sec:data_creation_mechan}, is used.

\begin{figure}[!h]
	\centering
	\includegraphics[width=0.9\linewidth]{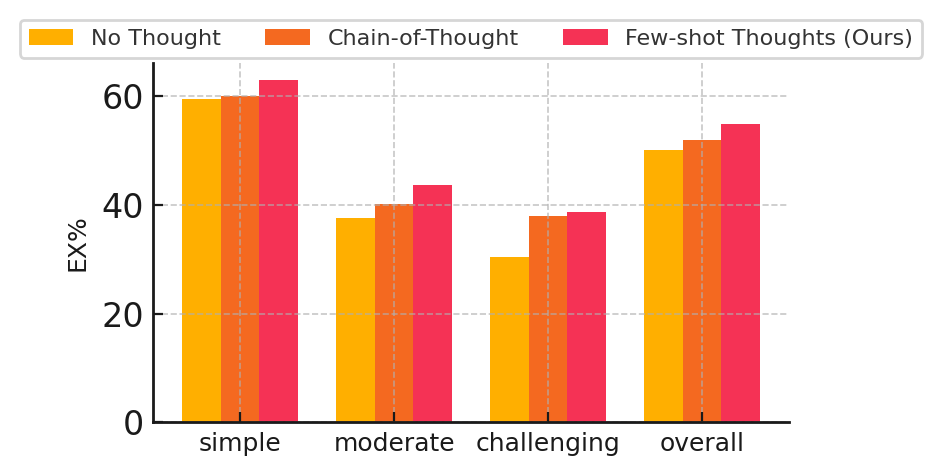}
	\vspace{-1em}
	\caption{Planner prompt comparison on BIRD Dev.}
	\label{fig:planner_prompt_comparison}
\end{figure}

The results in \autoref{fig:planner_prompt_comparison} show that Few-shot
Thoughts (3) consistently outperforms CoT and No Thought. In simple queries,
adding thought-based prompting provides only a slight improvement, with
execution accuracy increasing from 59.46\% (1) to 60.11\% (2) and 63.03\%
(3). However, for more complex queries, the impact is more significant. In
moderate queries, Few-shot Thoughts achieves 43.75\%, improving by 6.25\%
over No Thought. In challenging queries, Few-shot Thoughts (3) reaches
38.62\%, compared to 37.93\% of CoT and 30.34\% of No Thought, highlighting
the importance of well-structured thought processes for SQL generation.

\subsection{Database Domain Adaption}
\label{sec:database_domain_adaption}

\begin{figure}[t]
	\centering
	\includegraphics[width=0.9\linewidth]{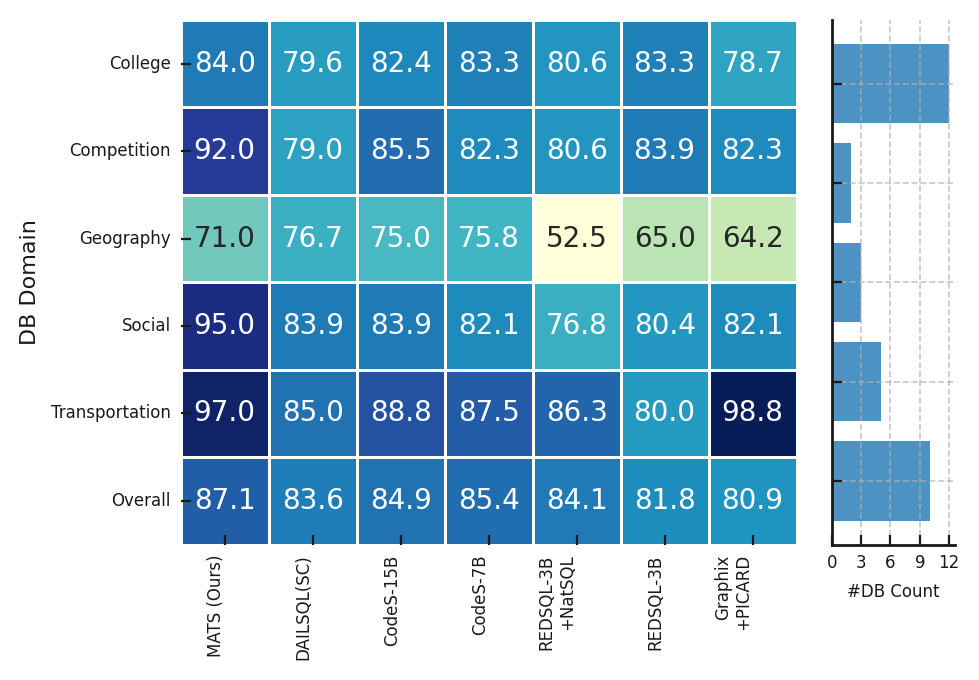}
	\vspace{-1em}
	\caption{EX\% comparison on different domains on Spider.}
	\label{fig:domain_knowledge}
\end{figure}

As shown in \autoref{fig:domain_knowledge}, our method \toolname, achieves the
best overall EX\% across database domains in the Spider benchmark. While
\toolname performs better than existing baselines such as DAILSQL(SC) and
REDSQL-3B + NatSQL in domains like \textit{Competition} and \textit{Social},
its performance is slightly lower in certain domains such as
\textit{Geography}. This may be due to larger models using in DAIL-SQL and
CodeS-7B, Codes-15B having better background knowledge in Geography. Despite
these variations, \toolname demonstrates strong generalization across different
domains.

\begin{figure}[t]
	\centering
	\includegraphics[width=0.9\linewidth]{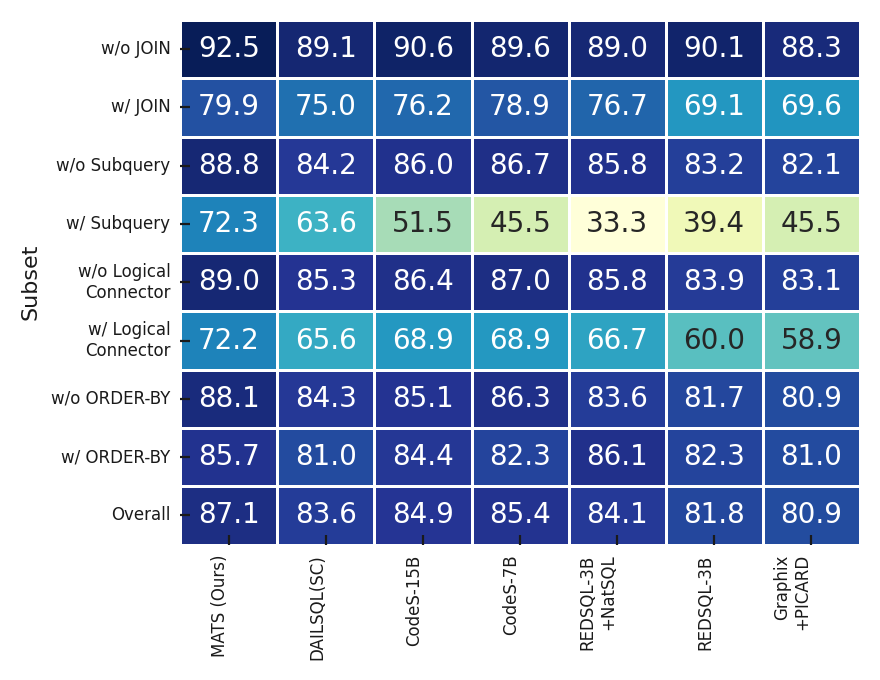}
	\vspace{-1em}
	\caption{EX\% comparison on different SQL characteristics on Spider.}
	\label{fig:sql_characteristic_spider}
\end{figure}

\begin{figure}[t]
	\centering
	\includegraphics[width=0.9\linewidth]{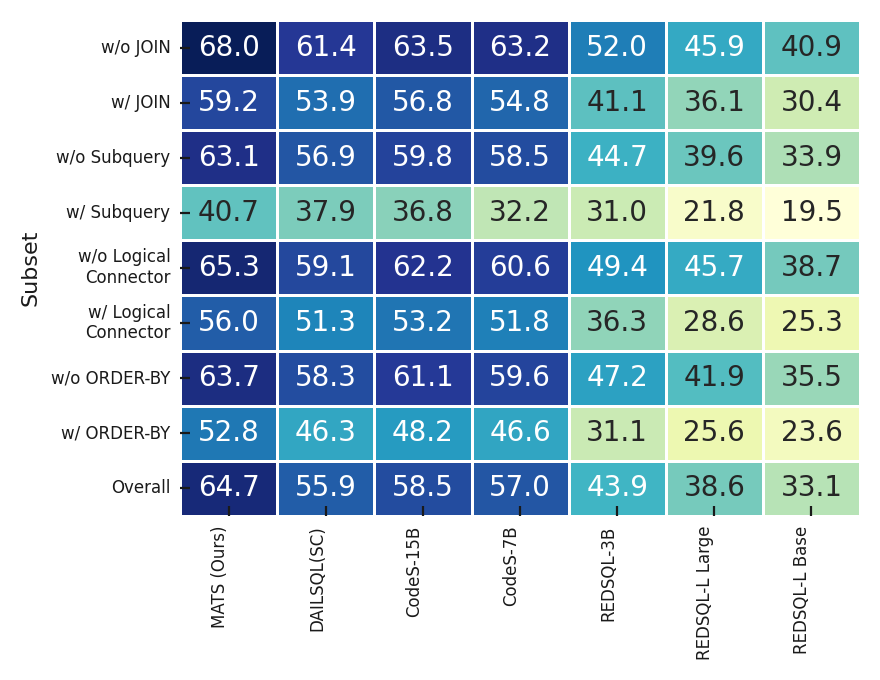}
	\vspace{-1em}
	\caption{EX\% comparison on different SQL characteristics on BIRD.}
	\label{fig:sql_characteristic_BIRD}
\end{figure}

\subsection{Evaluating on SQL Characteristics}

\autoref{fig:sql_characteristic_spider}
and \autoref{fig:sql_characteristic_BIRD} compare EX\% across SQL
characteristics on the Spider and BIRD datasets. \toolname (Ours) consistently
surpasses baselines like DAILSQL(SC), CodeS-15B, and REDSQL-3B, especially
on simpler queries without JOINs, ORDER-BY, or logical connectors. However,
subqueries remain a major challenge for all methods, with EX\% dropping
10-30\% compared to non-nested queries. While \toolname outperforms other
approaches in this aspect, it still experiences a noticeable decline,
indicating the inherent difficulty of handling complex SQL structures.

\section{Related Work}
\label{sec:related}

While we already discussed existing methods for Text2SQL, we here focus on
work that is related to the underlying techniques of \toolname, i.e., for
reinforcement learning, preference alignment, and the agent collaboration.

\sstitle{Reinforcement Learning with Human Feedback}
Reinforcement Learning with Human Feedback (RLHF) is a widely used approach for aligning language models with human preferences by integrating human evaluations directly into the training process. In RLHF, a reward model is trained on human-labeled data to score outputs based on their alignment with user intent. This reward signal is then used by reinforcement learning algorithms, such as Proximal Policy Optimization (PPO)~\cite{schulman2017proximal}, to optimize the model's behavior. RLHF has been instrumental in enabling language models to generate responses that are more aligned with human expectations and preferences. For example, this methodology has been successfully applied to instruction-following models like InstructGPT, significantly improving their capability to adhere to user guidelines while performing a broad spectrum of tasks~\cite{ouyang2022training}. This technique bridges the gap between general-purpose pretraining and task-specific utility, solidifying RLHF's role as a pivotal tool in modern AI systems.

Inspired by this paradigm, we propose a new reinforcement learning mechanism using execution feedback (RLEF) to further refine the \toolname agents. By using the responses of query execution, we save human efforts in providing labels like RLHF.

\sstitle{Alignment without Reward Models}
To overcome the complexities and instability associated with Reinforcement
Learning with Human Feedback (RLHF), particularly the challenges posed by
Proximal Policy Optimization (PPO)~\cite{schulman2017proximal}, alternative
alignment methods have been proposed.
Direct Policy Optimization (DPO) simplifies the alignment process by
integrating reward modelling directly into the preference learning stage,
eliminating the need for multi-stage training~\cite{rafailov2023direct}.
ORPO~\cite{hong-etal-2024-orpo} builds on DPO by integrating an odds
ratio-based penalty into supervised fine-tuning. ORPO combines SFT loss with
a relative ratio loss, streamlining training by removing the need for
separate reward models or alignment stages, thus enhancing efficiency and
scalability.

In this paper, we propose a new application of ORPO in Text2SQL tasks by imposing the loss on the completion part only. This makes the framework still light-weight while improving the accuracy of agents and simplifying the training process without requiring a separate reward model.

\sstitle{Multi-Agent Systems}
Recent advancements in multi-agent systems have demonstrated their
effectiveness in solving complex tasks by breaking them down into smaller,
specialized subtasks, each handled by an individual agent~\cite{wei2025enriching,peng2025stpe}. These systems
leverage the capabilities of Large Language Models to enable efficient
collaboration and communication among agents, leading to improved task
accuracy and efficiency.
Multi-agent frameworks have been successfully
applied across various domains, such as software engineering, where tools
like ChatDev~\cite{qian2024chatdev} facilitate collaborative coding and
debugging, and societal simulations, where MAS models human interactions to
evaluate policy outcomes~\cite{park2023generative}. Moreover, multi-agent
systems have
also been applied in recommendation systems, where agents work
collaboratively to
process user preferences, filter content, and generate personalized
recommendations through effective coordination and task
allocation~\cite{zhang2024agentcf}.

Inspired by these advancements, our framework adopts a multi-agent approach
for Text2SQL tasks, assigning dedicated agents for schema filtering, query
generation, validation, and error correction. This collaborative structure
significantly enhances SQL generation accuracy, especially for small
language models, while improving scalability through reinforcement learning
techniques.
\section{Conclusion}
\label{sec:con}

In this work, we introduced \toolname, a novel multi-agent framework designed for
Text2SQL tasks. The framework is optimized for small language models. Unlike
traditional LLM-based approaches, \toolname efficiently decomposes SQL generation
into specialized agent roles, allowing for effective execution in
resource-constrained environments. That is, agents filter irrelevant schema
elements and retrieve relevant column values; they generate multiple SQL
queries; they evaluate their outputs using database
responses; they refines SQL queries; and they eventually select the best SQL
query from a set of candidates.
Moreover, to further enhance performance, we proposed Reinforcement Learning
with
Execution Feedback, which improves SQL generation by leveraging
execution results as feedback to guide agent alignment.
Through extensive
evaluations on the BIRD and Spider datasets, we demonstrated that \toolname
achieves execution accuracy comparable to larger, closed-source LLMs while
being deployable on a single 24GB GPU. Our results indicate that \toolname
outperforms other open-source models in execution accuracy and robustness,
narrowing the gap between closed-sourced solutions that employ proprietary,
external models.

Despite these advancements, challenges remain in further improving
accuracy. Future work shall explore enhancing agent collaboration and
refining selection strategies to maximize execution accuracy. We can also improve the response of the database system itself to increase the quality of execution feedback, help the agents to learn faster and better. Additionally,
we will integrate broader domain-specific knowledge to further improve SQL
generation for real-world applications.

\balance



\end{document}